\documentclass[prb,twocolumn,superscriptaddress,showpacs,aps]{revtex4-1}
\pdfoutput=1

\usepackage{amsmath}
\usepackage{amstext}
\usepackage{graphicx}
\usepackage{rotating}
\usepackage{tabularx}
\usepackage[english]{babel}

\begin{document}

\title{Electronic Structure and Optical Properties of the Lonsdaleite Phase of Si, Ge and diamond}

\author{Amrit De}
\affiliation{Department of Physics and Astronomy and Optical Science and Technology Center, University of Iowa, Iowa City, Iowa 52242}
\affiliation{Department of Physics and Astronomy, University of California, Riverside, CA 92506}
\author{Craig E. Pryor}
\affiliation{Department of Physics and Astronomy and Optical Science and Technology Center, University of Iowa, Iowa City, Iowa 52242}

\date{\today}
\begin{abstract}
Crystalline semiconductors may exist in different polytypic phases with significantly different electronic and optical properties. In this paper, we calculate the electronic structure and optical properties of diamond, Si and Ge in the lonsdaleite (hexagonal-diamond) phase. We use an empirical pseudopotentials method based on transferable model potentials, including spin-orbit interactions. We obtain  band structures, densities of states and complex dielectric functions calculated in the dipole approximation for light polarized perpendicular and parallel to the $c$-axis of the crystal. We find strong polarization dependent optical anisotropy. Simple analytical expressions are provided for the dispersion relations. We find that in the lonsdaleite phase, diamond and Si remain indirect gap semiconductors while Ge is transformed into a direct gap semiconductor with a significantly smaller band gap.
\end{abstract}

\maketitle
\section{Introduction}

It is well known that under extreme  conditions a crystalline material may undergo a structural transition to a phase which remains effectively stable upon returning to standard thermodynamic conditions. The existence of such phases, polymorphism in compounds and allotropy in elemental crystals, results in materials with differing electrical and optical properties, such  diamond and graphite.
Polytypism is a particular case of polymorphism in which the coordination number does not change.
For example, III-V and group-IV semiconductors can crystallize in their non-naturally occurring polytypic phase while maintaining their tetrahedral coordination\cite{Mujica2003}.
For example, III-V semiconductors can crystalize in either the  zincblende (ZB) phase or the hexagonal wurtzite (WZ) phase.
Under extreme conditions cubic diamond (CD) transforms to a hexagonal wurzite crystal,  known as lonsdaleite (LD).
The form of diamond was not discovered until  1967 when it was found in a meteorite\cite{Frondel1967}.
It was synthesized soon thereafter\cite{Bundy1967} and is now believed to be the hardest known substance\cite{Pan.prl.2009}.

The high pressure phases of Si and Ge have been investigated\cite{Ackland2001,Mujica2003} and it has been determined that both undergo a series of structural phase transitions from  cubic, to $\beta$-Sn, to simple hexagonal,
to an orthohombic phase, to hexagonal closed packed,
to a face centered cubic  phase\cite{Ackland2001,Mujica2003}.
On the other hand, III-V semiconductors typically  crystalize either in $\beta$-Sn, nickeline (NiAs) or rocksalt structures when subjected to high temperature and pressure\cite{Ackland2001,Mujica2003}. More recently, the growth of WZ phase bulk GaAs was achieved under extreme conditions \cite{McMahon2005}.

Extreme temperatures and pressure are not the only ways to achieve WZ or LD growth.
Laser ablation have been used to synthesize stable LD phase Si \cite{Zhang1999}
III-V nanowires tend to crystalize in the WZ phase\cite{Koguchi1992JJAP,Mattila2006,Tomioka2007JJAP}.
This has been attributed to various factors such as the small nanowire radii \cite{Akiyama2006PRB,Galicka2008JP}, growth kinetics\cite{Dubrovskii2008PRB}, interface energies\cite{Glas2007} and electron accumulation at the catalyst's interstitial site \cite{Haneda2008}.
It is now believed that the tendency of nanowires to crystalize in WZ/LD phase may be true for group-IV semiconductors as well\cite{Arbiol2008}.
It has been experimentally found that Si nanowires with a radius in excess of 10 nm tend to crystalize in the LD phase \cite{Fontcuberta2007} and a number of recent theoretical investigations have confirmed that the hexagonal LD phase is the more stable for Si nanowires exceeding certain critical dimensions\cite{Ponomareva2005,Kagimura2005,Zhang2008,Liu2009}.
Similar structural phase transitions are expected for Ge nanowires as well \cite{Kagimura2005}.

Semiconductor nanowires  have attracted much interest due to their potential  applications such as photovoltaic cells\cite{Hu2007,Tsakalakos2007,Czaban2009}, nano-mechanical resonator arrays \cite{Henry2007}, THz detectors \cite{Balocco2005,Gustavsson2008}, single photon detectors\cite{Rosfjord2006,Zinoni2007,Dauler2007},  field-effect transistors\cite{Cui2003,Greytak2004}, single-electron transistors, and other devices\cite{Thelander2003,Bjork2002,Bjork2002b,Panev2003,Duan2001,Samuelson2004}.
In addition, the quasi 1-D nature of  nanowires  allows materials with large lattice mismatches to be combined to form hetrostructures that are not possible in planar structures.
The design and characterization of such devices requires an understanding of the electronic and optical properties of WZ/LD phase semiconductors.
Even though bulk LD phase Si, Ge and diamond have been synthesized in the laboratory, their electronic structure still remains experimentally unverified.

Since these materials may be used for spin based devices,  spin-orbit interactions should be included.
The band structures and dielectric functions for LD phase diamond, Si and Ge has been calculated  using empirical pseudopotentials \cite{Joannopoulos1973}, however these calculations were done without the inclusion of spin-orbit interactions.
The bandstucture of LD phase diamond Si and Ge have also been calculated using density functional theory \cite{Salehpour1990,Murayama1994,Raffy2002}, however these methods  have well known shortcomings in predicting energies of excited states.
Therefore, there is a need for accurate empirically based band structures including spin.

In this paper, we first present bulk electronic band structure calculations for diamond, Si and Ge in the LD phase using empirical pseudopotentials with the inclusion of spin-orbit interactions. These calculations are based on transferable model pseudopotentials assuming an ideal LD structure. The spherically symmetric ionic model potentials are first obtained by fitting the calculated bulk energies of the cubic polytype to experiment. The band structure of the LD polytype is then obtained by transferring the spherically symmetric model pseudopotentials to the LD crystal structure. These model potentials are expected to be transferable between polytypes due to the similarities in their crystal structures.  Like the cubic structure, the LD structure is built from tetrahedrons of the same atom but are stacked differently.
In both structures all of the nearest neighbors and nine out of the twelve second nearest neighbors are at identical crystallographic locations \cite{Birman1959} and all the second nearest neighbors are equidistant. Hence the local electronic environment should be very similar in both polytypes.
This method has proven to be quite successful in obtaining the bulk band structures of various semiconductor polytypes in the past\cite{Bergstresser1967,Joannopoulos1973,Foley1986,Bandic1995,Pugh1999,Pennington2001,Fritsch2003,Fritsch2006,Cohen.book}. We have used this method to predict the band structures of WZ phase III-V semiconductors\cite{De2010prb} and our calculations are in excellent agreement with experiment for the cases for which the WZ band gaps are known (namely GaAs, InP and InAs). Since then, a number of recent experiments have  provided further conformations of the predicted band gaps (and their respective symmetries) \cite{Novikov2010,Ketterer2011,Heiss2011,Peng2012,Jahn2012} and the  effective mass of InAs$_{wz}$\cite{Wallentin2011}.



 We have also calculated the  dielectric functions for diamond, Si and Ge in the LD phase in the linear response regime within the electric dipole approximation for light polarized parallel and perpendicular to the c-axis of the crystal.
These calculations are in part motivated by recent experimental results showing that the photoluminescence(PL) intensity in nanowires is strongly polarization dependent \cite{Wang2001,Mattila2006,Mattila2007,Mishra2007,Lan2008,Kobayashi2008,Novikov2010}, including Si nanowires\cite{Ma2005}.
Nanowire polarization effects can arise from the dielectric mismatch at the wire surface \cite{Wang2001},  electron-hole recombination selection rules \cite{Birman1959b,Streitwolf1969,Sercel:1990p646}, as well as from the underlying WZ-type crystal structure\cite{Cardona1965,Reynolds1965,Xu1993,Ninomiya1995,Kawashima1997,Alemu1998}.
The dominant  effect could be more easily identified if the optical dielectric functions for these semiconductors were easily available. However, experimentally measuring optical dielectric functions for   LD phase semiconductors is difficult due to the extreme conditions required for growing bulk samples.


Our dielectric function calculations are carried out within the one-particle picture. Much has been done in recent years to include the effects of two particle contributions (such as electron-hole interactions) in the dielectric function  \cite{Rohlfing1998,Arnaud.prb.2001,Onida.rmp.2002,Laskowski.prb.2005,IsmailBeigi2008,Puangmali2008,Malone2010}. In the case of {\it first principles} calculations, this improves agreement with experiment for the low energy part of the optical spectrum  \cite{Rohlfing1998}. However, the dominant contribution to the dielectric function comes from one-particle calculations, which by itself tend to be in good agreement with experiment\cite{Trani2005,Wu2007,Makinistian2010}. Moreover, while two-particle corrections   improve {\it ab-initio} dielectric functions\cite{Cohen.book}, empirical methods include some two-particle effects through their fit to the experimental data (which includes such effects). Typically, dielectric functions calculated using EPMs are in good agreement with experiment \cite{Cohen.book}.

The required momentum matrix elements, required for our calculations, are obtained from the pseudopotential wave functions. In general, momentum matrix elements for pseudopotential calculations need to be corrected for the missing the core states. Such corrections are typically done using nonlocal terms \cite{Kageshima1997,Adolph2001,Pickard2001,Monachesi2001}, which can also account the exchange and correlations effects as well\cite{Gavrilenko1996,Gavrilenko1997,Schmidt2003}.
In our calculations, nonlocal effects are included in the form of spin orbit interactions only.

Another problem with {\it ab-initio} dielectric function calculations is that $\epsilon_0$ is often overestimated\cite{Hybertsen1987,Gironcoli1989,Levine1991}. This is then improved upon by the inclusion of local field effects \cite{Arnaud.prb.2001} such as electron-phonon interactions, which typically  affect the low frequency part of the dielectric function (terahertz regime). However, this is not necessary in our case since local field effects shift the peak positions\cite{Onida.rmp.2002}. The EPM includes such effects through the fitting to experiments which necessarily include the effects.

We instead adopt a simpler approach to take the missing core states and local field effects into account. We correct the static dielectric function, $\epsilon_0$, for the unknown polytype by making use of the {\it known} $\epsilon_0$ of the cubic phases of diamond, Si and Ge. First, the optical dielectric functions for these cubic group-IV semiconductors are calculated. The optical sum rules are then used to obtain a set of constants which normalizes the calculated cubic $\epsilon_0$ to their respectively known experimental values. Since the constituent element of each polytype is the same, corrections to account for the missing core states should be nearly the same and transferable between polytypes. These normalization constants are then used to correct $\epsilon_0$ for the unknown polytypes(LD), which therefore also corrects the LD phase dielectric functions as well. Unlike calculations involving local field effects, our simple approach fixes $\epsilon_0$ without shifting peak positions.

This paper is organized as follows.  Section \ref{sec:EPM}  outlines the empirical pseudopotential method, followed by a description of  both the cubic  and  hexagonal band structures, and a discussion of the respective symmetries of the two polytypes in section \ref{sec:Bands}. Our results are  in section \ref{sec:LD} where we present the calculated band structures, densities of states (DOS),  effective masses for zone center states and transition energies at various high symmetry points . The optical dielectric functions and reflectivity spectra are given in section \ref{sec:optical}. Finally, we summarize our results in section \ref{sec:summary}.

\section{Empirical Model Pseudopotentials}\label{sec:EPM}

We use the empirical pseudopotential formalism of Cohen and Chelikowsky \cite{Cohen.book}.
However, rather than  discrete form factors we  use continuous model potentials so they are transferable between polytypes. The pseudopotential Hamiltonian consists of the kinetic, local pseudopotential ($V_{pp}$) and SO interaction ($V_{so}$) terms,
\begin{eqnarray}
{H} = \frac{-\hbar^2\mathbf{K}^2}{2m} + {V}_{pp} + {V}_{so}.
\label{HoVso}
\end{eqnarray}
 In a periodic crystal $V_{pp}$ can be expanded in terms of plane waves as
\begin{eqnarray}
 V_{pp}({\mathbf r}) &=& \frac{1}{N}\displaystyle \sum_{{\bf G},\alpha}\sum_{j=1}^{N}V_\alpha^{FF} ({\bf G})e^{i\mathbf{G}\cdot (r-\boldsymbol \tau_{\alpha,j})}
 \label{EMP2}
\end{eqnarray}
where $\mathbf G$ are reciprocal lattice vectors,  $\alpha$ labels the atom type, $V^{FF}_\alpha({\bf G}) $ is the form factor of the $\alpha^{th}$ type of atom, $N$ is the number of atoms per unit cell of a given type, and $\boldsymbol  \tau_{\alpha,i}$ is the position of atom number $j$ of type $\alpha$.

For a compound with only one type of atom, the pseudopotential  is simply
\begin{eqnarray}
\langle\mathbf{G'}|{V}_{pp}|\mathbf{G}\rangle &=& V^{FF}(\mathbf{G'}-\mathbf{G})S(\mathbf{G'}-\mathbf{G})
\label{Ho.EPM}
\end{eqnarray}
where the structure factor is
\begin{eqnarray}
S(\mathbf{G})&=&\frac{1}{N}\displaystyle\sum_{j}\exp(-i\mathbf{G}\cdot{\boldsymbol\tau}_j) \label{SF}.
\end{eqnarray}

$V^{FF}$ can be obtained in various ways \cite{Cohen.book,Heine.book}. In the empirical pseudopotential approach the atomic form factors are adjusted so the  calculated energies at various high symmetry points  fit experiment. In order for pseudopotentials to be transferable between polytypes (having different $\bf G$'s)$V^{FF}({\bf G})$ should be a continuous function of $\bf G$. A wide variety of such model potentials have been used in literature\cite{Heine.book,Xia1988,Yeh1994b,Pugh1999,Fan2006}, and use potentials of the form
\begin{eqnarray}
V^{FF}({\bf G})&=&(x_1G + x_2)\left[1+\exp({x_3G^2 + x_4})\right]^{-1}~~~~~~~ \label{FF.func1}
\end{eqnarray}
where $G=|{\mathbf G}|\times a/2\pi$, and $x_{j}$ are adjustable parameters used to fit each material's cubic phase bandstructure to experiment. Model potentials that yield an accurate band structure of a known polytype should reliably predict the band structure for the unknown polytype if the electronic environment in the two crystals structures are similar.


 We take spin-orbit coupling into account by including the interaction\cite{Weisz1966,Chelikowsky1976}
\begin{eqnarray}
   \langle \mathbf{K'},s'| {V}_{so}|\mathbf{K},s\rangle&=&(\mathbf{K'}\times\mathbf{K})\cdot\langle{s'}|{\boldsymbol\sigma}|{s}\rangle \nonumber \\
   && \displaystyle\sum_{l} \lambda_l ~{\rm\it{P}}_l'(\cos\theta_{\mathbf { K'\cdot K} })S(\bf
   K'-K) \nonumber \\
\label{SOC1}
\end{eqnarray}
It is not necessary to expand Eq.~\ref{SOC1} beyond $l=2$ since group-IV semiconductors do not have core shells filled beyond $d$-orbitals.
For Ge the terms up to $l=2$ in Eq.~\ref{Vso.EMP1} are included, while for Si and diamond we  only go up to $l=1$.
Expanding Eq.~\ref{SOC1} up to $l=2$, the spin-orbit coupling term is
\begin{eqnarray}
& \langle \mathbf{K'},s'| {V}_{so}|\mathbf{K},s\rangle=
-i(\mathbf{\hat{K}'}\times\mathbf{\hat{K}})\cdot\langle{s'}| {\boldsymbol \sigma}|{s}\rangle \nonumber \\
&\left[       \left( \lambda_p+ \lambda_d \, \mathbf{\hat{K}'}\cdot\mathbf{\hat{K}} \right)  S(\mathbf{G'}-\mathbf{G})  \right]
\label{Vso.EMP1}
\end{eqnarray}
$\lambda_l$ is a coefficient that can be written in terms of the core wave functions
\begin{equation}
\lambda_l = \mu_l\beta_{nl}({\mathbf K'})\beta_{nl}({\mathbf K})
\label{lam}
\end{equation}
\begin{equation}
\beta_{nl}(K) = C
{\displaystyle{\int_0^\infty}}i^l\sqrt{4\pi(2l+1)}j_{nl}(Kr)R_{nl}(r)r^2dr
\label{beta}
\end{equation}
where $\boldsymbol \sigma$s are the Pauli matrices, $\bf{K=G+k}$,  $\theta$ is the angle between ${\bf K}$ and ${\bf K}'$, and $\mu_l$ are adjustable parameters used to fit the spin-orbit splitting energies to experiment \cite{Chelikowsky1976}.
The overlap integral, $\beta_{nl}$, is constructed from the radial part of the core wave function, $R_{nl}$, which is an approximate solution to the Hartree-Fock equations which we obtain from Herman-Skillman tables\cite{Herm.Skill}.
$C$ is a normalization constant such that $\beta_{nl}(K)/K$ approaches unity in the limit $K$ goes to zero.
Spin-orbit interactions are included for only the outer most $p$ shells ($n=4$ in Ge, $n=3$ in Si, and $n=2$ in diamond) and $d$ shells ($n=3$ in Ge).

\section{Band Structures and Crystal Symmetries} \label{sec:Bands}

\subsection{Lonsdaleite Crystal Structure and Symmetries} \label{sec:crystalStructure}

The LD/WZ crystal structure is constructed from two interpenetrating hexagonal close-packed (HCP) lattices, just as the diamond/ZB structure is constructed from two interpenetrating FCC lattices.
For ideal crystals, the lonsdaleite lattice constant is related to the diamond lattice constant as $a=a_{cubic}/ \sqrt{2}$ and the lattice constant along the $c$-axis (the [111] direction) is given by $c=\sqrt{1/u}~a$.
We assume an ideal LD crystal with $u=3/8$  in this paper, giving $a=2.522~\AA$ and $c=4.119~\AA$ for diamond, both of which are in  agreement  with the experimental values of $a=2.52~\AA$ and $c=4.12~\AA$ \cite{Bundy1967}.
In the case of LD phase Si, a wide range of measured lattice constants have been reported with $(a,c)=(3.84~\AA,6.280~\AA)$\cite{Zhang1999}, $(3.84~\AA,6.180~\AA)$\cite{Hendriks1984}, $(3.837~\AA,6.316~\AA)$\cite{Besson1987}.  Assuming an ideal LD crystal structure, we use $a=3.836~\AA$ and $c=6.264~\AA$ for Si. For Ge the ideal lattice constants are $a=3.993~\AA$ and $c=6.520~\AA$, which is within $1~\%$ of the experimental values\cite{Xiao1992} of $a=3.96~\AA$ and $c=6.57~\AA$.

When viewed along the [111] direction, the inter-layer atomic bonds in LD lie in an eclipsed conformation, defining the axis of hexagonal symmetry while in the CD structure the inter-layer atomic bonds are in a staggered conformation, making all four body diagonals of the cube  equivalent. The nearest neighbors are  the same in the two polytypes (due to their tetrahedral symmetry), and nine of the twelve next nearest neighbors are in identical positions in both crystals and the remaining three next nearest neighbors are equidistant. These structural similarities suggests that the local electronic environment will be very similar in the two crystals, and hence the atomic form factors should be nearly identical in both polytypes.

The LD crystal structure has a space group symmetry classification of $D_{6h}^4$(or $P6_3/mmc$). It has inversion symmetry in addition to all the symmetries of WZ. The irreducible representation of the space group of $\Gamma$ are just the representations of the point group $D_{6h}$ (which has all the symmetries of $C_{6v}$ as well as inversion symmetry). While moving along the $k_z$ direction the symmetry is lowered to $C_{6v}$. The $A$, $K$ and $H$  all have point group symmetries of $D_{3h}$ \cite{Dresselhaus.book}. $D_{3h}$ is isomorphic to $C_{6v}$ and is a symmorphic invariant subgroup of $D_{6h}$. The $L$ and $M$ points have $D_{2h}$ symmetry.
The point group operations must be followed by appropriate translations in order to obtain the irreducible representations of the wave functions at the high symmetry points. For example at the $\Gamma$ point the point group operations must be followed by a translation $\tau = (0,0,c/2)$.

The LD crystal structure has lower  symmetry than the cubic diamond structure, and the  SO interaction leads to additional lifting of orbital degeneracies. In the absence of spin-orbit coupling, the hexagonal crystal field of LD splits the $p$-like $\Gamma_{15}$ state of cubic structure into a four-fold degenerate  $\Gamma_6 $ and a doubly degenerate $\Gamma_1$. In terms of the $p$-orbitals, these states are $p_z\rightarrow\Gamma_1$ are $p_x,p_y\rightarrow\Gamma_6$. With the inclusion of spin-orbit coupling, $\Gamma_{6v}$ splits into the $\Gamma_{9v}$ heavy-hole and the $\Gamma_{7v}$ light-hole. Therefore, all zone center states in LD belong to either $\Gamma_7$, $\Gamma_8$, or $\Gamma_9$ with either even or odd parity since LD has a center of inversion.
Unlike WZ, there are no spin splittings in LD due to its inversion symmetry.
%

                                \subsection{Predicted Lonsdaleite Band Structures}\label{sec:LD}

We use spherically symmetric local form factors  fit to the CD band structure, which are then transferred to the LD crystal. The potentials should be transferable since the local electronic environments of the two polytypes are very similar.
The lonsdaleite primitive unit cell has four atoms. We choose the origin, so that the atoms are located at
${\bf v}_1 = \frac{1}{3} {\bf a_1}+\frac{2}{3} {\bf a_2} $,
${\bf v}_2 = \frac{2}{3} {\bf a_1}+\frac{1}{3} {\bf a_2} + \frac{1}{2} {\bf a_3}$,
${\bf v}_3 = {\bf v}_1 + u{\bf a_3}$ and
${\bf v}_4 = {\bf v}_3 + u{\bf a_3}$. Where ${\bf a}_1=(1,0,0) a$, ${\bf a}_2=(-1,\sqrt{3},0) a/2$ and ${\bf a}_3=(0,0,c)$ are the the primitive lattice vectors. Substituting these atomic positions into Eq.~\ref{SF}, we obtain the following structure factor
\begin{eqnarray}
S &=& \frac{1}{2} \exp\left(  -\frac{iG_ya}{\sqrt{3}} - \frac{iG_z uc}{2} \right)\cos\left( \frac{G_zuc}{2} \right)\times~~~~~~\\\nonumber
&~&\left[1+ \exp\left( -\frac{iG_xa}{2} + \frac{iG_ya}{2\sqrt{3}} - \frac{iG_zc}{2} \right) \right]
\label{SF.WZ}
\end{eqnarray}
where, $G_j$ ($j=x,y,z$) are the components of the reciprocal lattice vector $\bf{G}$.


\begin{figure}\center
  \includegraphics[width=1\columnwidth]{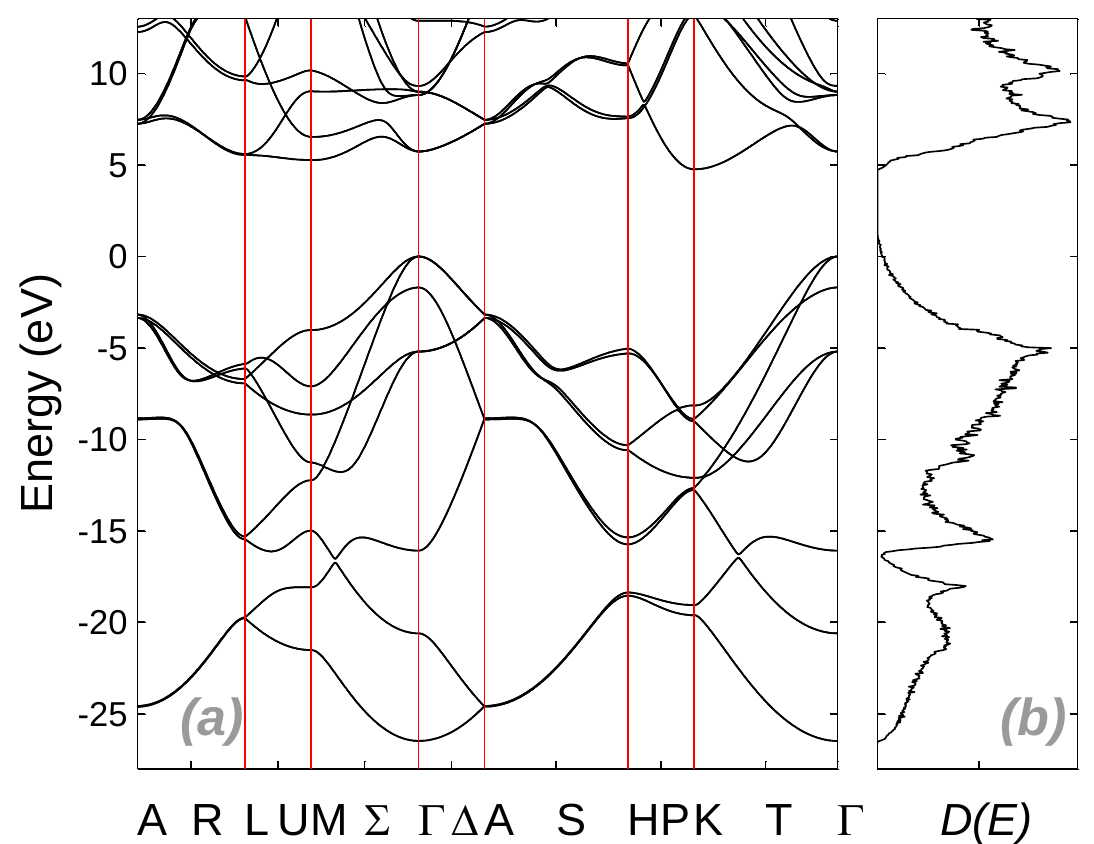}
  \caption{ ({\bf a}) Band structure for diamond in the lonsdaleite phase ({\bf b}) Density of states.}
  \label{fig:WZ-C}
\end{figure}

\begin{figure}\center
  \includegraphics[width=1\columnwidth]{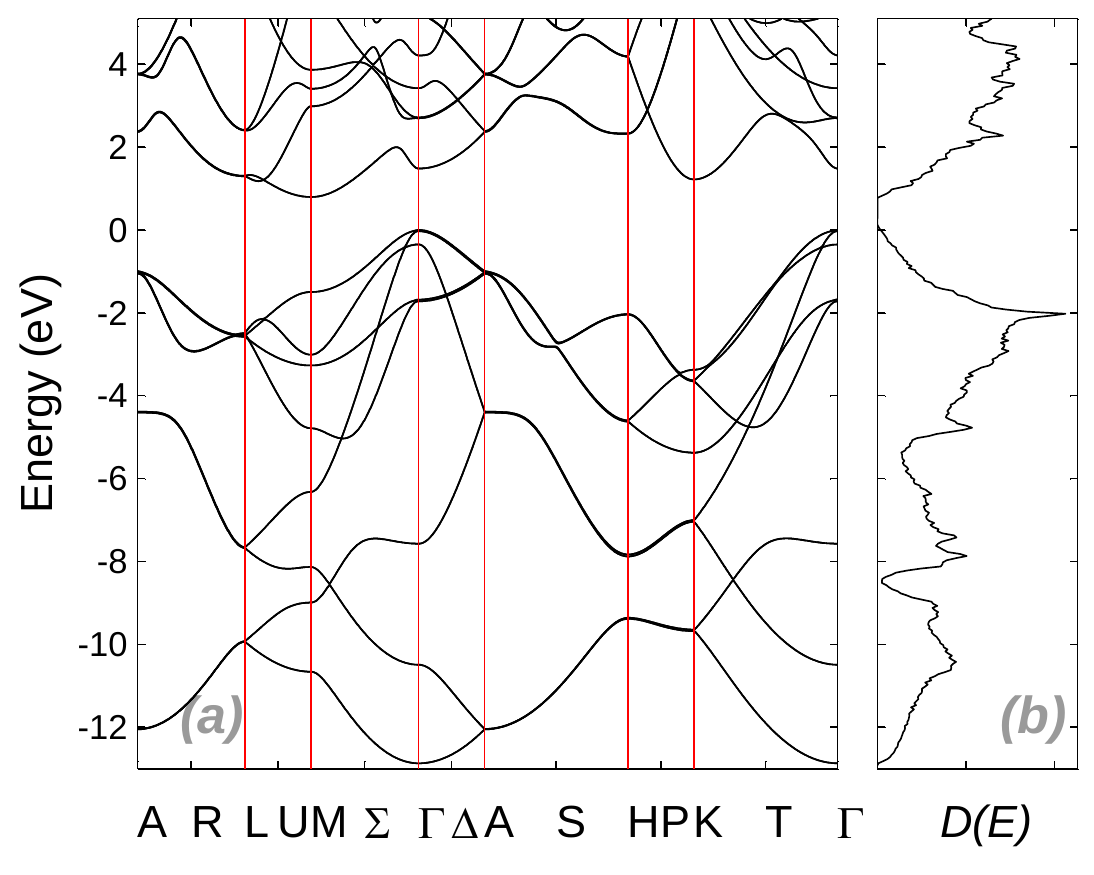}
  \caption{ ({\bf a}) Band structure for Si in the lonsdaleite phase. ({\bf b}) Density of states.}
  \label{fig:WZ-Si}
\end{figure}

\begin{figure}\center
  \includegraphics[width=1\columnwidth]{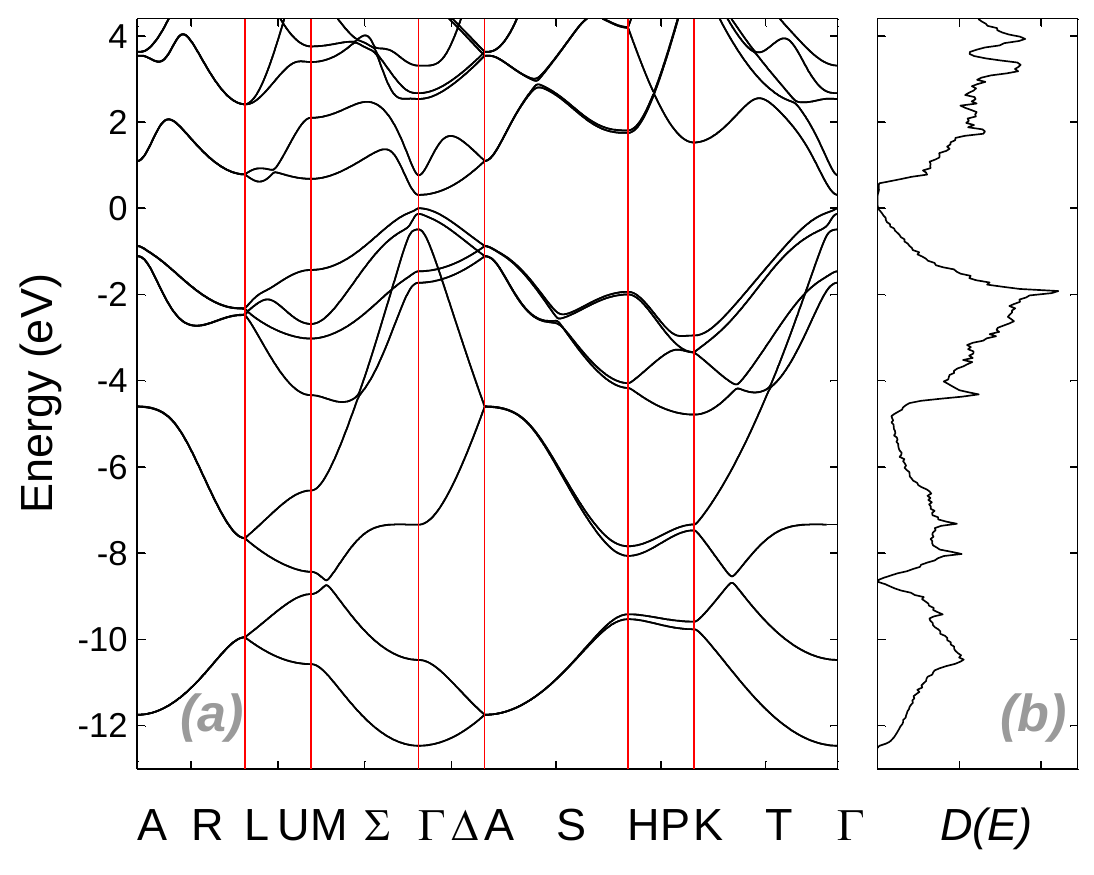}
  \caption{ ({\bf a}) Band structure for Ge in the lonsdaleite phase. ({\bf b}) Density of states.}
  \label{fig:WZ-Ge}
\end{figure}

The calculated band structure and the corresponding density of states(DOS) for diamond, Si and Ge in LD phase are shown in Figs. \ref{fig:WZ-C}-\ref{fig:WZ-Ge}. The electronic band structures are calculated in the irreducible wedge of the Brillouin zone. The LD band structure  is more complicated its CD counter part due to its lower crystal symmetry. For a given energy range, there are roughly twice as many bands for the LD phase. The points, $A$ and $H$, are special points where the energy levels stick together because the structure factor is zero there.

Due to the similarities of the two crystals, many of  the high symmetry points in the Brillouin zones of cubic and LD have a one-to-one correspondence with each other (just as in the case of ZB and WZ). This one to one correspondence is particularly useful in interpreting their respective band structures.

The volume of LD's first Brillouin zone is about half of that of its cubic counter part. Therefore, if one were to take an intersection of the two Brillouin zones such that each of their $\Gamma$-points conincide, then the $L$-point in the cubic structure also coincides with the $\Gamma$-point in LD \cite{Murayama1994}. Thus, in the free electron model, the zone center LD/WZ energies can be directly predicted from the cubic $\Gamma$ and $L$ point energies. In single group notation (absence of SO interactions), the cubic $\Gamma_1$, $L_1$ and $L_3$ states correspond to $\Gamma_1$, $\Gamma_3$ and $\Gamma_5$ respectively in LD. However, the presence of the crystal potential will perturb the exact one to one correspondence of the high symmetry point energies. Because of this zone folding of the $L$-valley, indirect gap cubic materials with an $L$ valley conduction band minima could be expected to have a direct gap in the LD phase (unless the energy of that state was significantly shifted by the crystal potential).

There also are similar correspondences between the high symmetry directions of the two crystals. The $\Lambda ( \Gamma\rightarrow L$) line in the cubic structure corresponds to the $\Delta(\Gamma \rightarrow A)$ line in the LD\cite{Birman1959} structure. Note that there are eight equivalent $L$ directions in the cubic structure. Only the ones that are along the $c-axis$ map on to the $\Gamma_{hex}$ point. The other six along with $X_{cub}$ maps on to a point on the $U_{hex}$-line, two thirds away from the $M_{hex}$ point. We label this point as $M'_{hex}$. The $\Delta_{cub}$ line maps onto a line joining $M'_{hex}$ and $\Gamma_{hex}$. The $\Delta_{cub}$ line is especially important as the band gap for both diamond and Si lies along this line close to $X_{cub}$.


We list the energies at various high symmetry points along with the corresponding irreducible representations of these states in tables \ref{tab:WZ-C}-\ref{tab:WZ-Ge}.The irreducible representations of the zone center states was determined by transforming the pseudo wave functions under the symmetry operations of the respective crystallographic point group. The zone center ($\Gamma$ point) effective masses for diamond, Si and Ge for ${\bf k}$ parallel and perpendicular to the $c$-axis is shown in table\ref{tab:WZ-mass}.

From the band structure calculations, it is seen that in the LD phase, diamond and Si are indirect gap semiconductors. Diamond has a band gap of about $4.767$eV with band minima occuring at the $K$-valley. This is in agreement with earlier results obtained by Salehpour and Sathpathy\cite{Salehpour1990} where the band minima was also shown to occur at $K$. Their estimated band gap from LDA calculations (with corrections) was about $4.5$ eV. The LD band gap for diamond is also significantly smaller than its cubic phase band gap of $5.4$ eV.

In the case of the LD phase Si, the calculated band gap is $0.796~\rm eV$ and the band minima is at the $M$-valley. The LD phase gap for Si is a lot smaller than its cubic badgap of about $1.1~\rm eV$.  In our calculations, Ge is the only direct gap group-IV semiconductor in the LD phase. It has a $\Gamma_{8}$ conduction band minima. The $\Gamma_{8}$ symmetry is due to the fact that cubic phase Ge has a conduction band minima at the $L$ valley, which folds over to $\Gamma_{8}$ in the hexagonal phase.

With the inclusion of SO interactions for LD, the top three valence states are typically  (in  order of decreasing energy) $\Gamma_9$, $\Gamma_7$, $\Gamma_7$ (normal ordering)\cite{Birman1959b,Thomas1959} or  $\Gamma_7$, $\Gamma_9$, $\Gamma_7$  (anomalous ordering) which results from a negative spin-orbit energy. In our calculations, the top three valance band states in Si and Ge have normal ordering in LD phase, while the top valance band states in diamond have neither normal nor anomalous ordering. Instead the top two valance band states for LD-diamond are $\Gamma_9^-$ and $\Gamma_8^-$. The $\Gamma_7^+$, $\Gamma_7^+$, $\Gamma_9^+$ states lie further bellow.

The spin-orbit splitting energy, $\Delta_{so}$ and the crystal field splitting, $\Delta_{cr}$, can be extracted using the quasi cubic approximation which assumes the WZ/LD structure to be equivalent to a $[111]$-strained zincblende structure \cite{BirPikus}. $\Delta_{so}$ and $\Delta_{cr}$  are related to the $\Gamma_{7v}^1$ light hole and $\Gamma_{7v}^2$ light hole energies by
\begin{eqnarray}
E(\Gamma_{7v}^{1,2}) - E(\Gamma_{9v}) &=& -\frac{\Delta_{so}+\Delta_{cr}}{2}\nonumber\\
&\pm&\frac{1}{2}\left((\Delta_{so}+\Delta_{cr})^2-\frac{\Delta_{so}\Delta_{cr}}{u}\right)^{\frac{1}{2}}~~~~~~~~~
\label{WZ:CrSo}
\end{eqnarray}
where  $\sqrt{u}=a/c$  is $3/8$ for an ideal WZ/LD structure assumed here. The band ordering and irreducible representations of the zone center states need to be identified before using this equation.
We have complied a shorter table (table-\ref{tab:WZ-compact}) for the LD phase semiconductors, that lists the band gap, $\Delta_{so}$, $\Delta_{cr}$, and the offset between the valence band edges of each polytypes.

\begin{table*}
\begin{tabular}{|c|c|c|c|c|c|c|c|c|c|c|c|}
\hline\hline
\textbf{IR}&\textbf{E (eV)}&\textbf{IR}&\textbf{E (eV)}&\textbf{IR}&\textbf{E (eV)}&\textbf{IR}&\textbf{E (eV)}&\textbf{IR}&\textbf{E (eV)}&\textbf{IR}&\textbf{E (eV)}\\\hline
$\Gamma_7^+$ & -26.4861 & $A_7$ & -24.6126 & $L_5$ & -19.7797 & $M_5$ & -21.5188 & $H_7$ & -18.5436 & $K_7$ & -19.6131\\
$\Gamma_8^-$ & -20.6056 & $A_8$ & -24.5697 & $L_5$ &   -19.73 & $M_5$ & -18.0857 & $H_8$ & -18.3647 & $K_8$ & -19.0582\\
$\Gamma_8^-$ &  -16.081 & $A_8$ &  -8.9149 & $L_5$ & -15.4592 & $M_5$ & -14.9875 & $H_8$ & -15.7359 & $K_8$ & -12.7624\\
$\Gamma_8^+$ &  -5.2053 & $A_7$ &  -8.8447 & $L_5$ &  -15.313 & $M_5$ & -12.2387 & $H_8$ & -15.3608 & $K_8$ & -12.6617\\
$\Gamma_9^+$ &  -5.1929 & $A_8$ &  -3.3676 & $L_5$ &  -6.9353 & $M_5$ & -11.2565 & $H_9$ & -10.5802 & $K_9$ & -12.0976\\
$\Gamma_7^+$ &  -1.6965 & $A_9$ &  -3.3601 & $L_5$ &  -6.6979 & $M_5$ &  -8.6464 & $H_7$ & -10.3175 & $K_8$ &  -9.0102\\
$\Gamma_7^-$ &   -0.003 & $A_9$ &  -3.1847 & $L_5$ &  -6.1246 & $M_5$ &   -7.101 & $H_8$ &   -5.303 & $K_9$ &  -8.8838\\
$\Gamma_9^-$ &        0 & $A_7$ &  -3.1772 & $L_5$ &  -5.8708 & $M_5$ &  -4.0212 & $H_9$ &    -5.03 & $K_7$ &  -8.1377\\
$\Gamma_7^-$ &   5.7365 & $A_7$ &   7.2555 & $L_5$ &   5.5528 & $M_5$ &   5.2688 & $H_9$ &   7.5599 & $K_9$ &   4.7676\\
$\Gamma_9^-$ &    5.741 & $A_9$ &   7.2574 & $L_5$ &   5.6029 & $M_5$ &   6.5429 & $H_7$ &   7.6419 & $K_7$ &  13.1856\\
$\Gamma_7^-$ &   8.8245 & $A_8$ &   7.4665 & $L_5$ &   9.6269 & $M_5$ &   9.0309 & $H_9$ &  10.4545 & $K_9$ &  13.4011\\
$\Gamma_9^-$ &   8.8246 & $A_9$ &   7.4685 & $L_5$ &   9.8392 & $M_5$ &  10.1738 & $H_7$ &  10.5574 & $K_7$ &  14.0873\\
$\Gamma_9^+$ &   9.0085 & $A_7$ &  12.2676 & $L_5$ &  13.1219 & $M_5$ &   18.184 & $H_9$ &  17.4962 & $K_9$ &   17.121\\
$\Gamma_8^+$ &   9.0095 & $A_8$ &  12.5631 & $L_5$ &  13.3848 & $M_5$ &  18.2476 & $H_8$ &  17.5338 & $K_8$ &  17.5896\\
$\Gamma_7^+$ &   9.3188 & $A_7$ &   15.361 & $L_5$ &   24.929 & $M_5$ &  20.3734 & $H_9$ &  19.3015 & $K_9$ &  18.8105\\
\hline
\end{tabular}
\caption{Transition energies at various high symmetry points and the respective irreducible representations(IR) for the lonsdaleite phase of diamond. Note that the conduction band minima is in near vicinity of the $K$ valley }
\label{tab:WZ-C}
\end{table*}

\begin{table*}
\begin{tabular}{|c|c|c|c|c|c|c|c|c|c|c|c|}
\hline\hline
\textbf{IR}&\textbf{E (eV)}&\textbf{IR}&\textbf{E (eV)}&\textbf{IR}&\textbf{E (eV)}&\textbf{IR}&\textbf{E (eV)}&\textbf{IR}&\textbf{E (eV)}&\textbf{IR}&\textbf{E (eV)}\\\hline
$\Gamma_7^+$ & -12.8713 & $A_7$ & -12.0498 & $L_5$ & -9.9375 & $M_5$ & -10.6599 & $H_7$ & -9.3824 & $K_7$ & -9.6803\\
$\Gamma_8^-$ & -10.4924 & $A_8$ & -12.0433 & $L_5$ & -9.9237 & $M_5$ &   -8.992 & $H_8$ & -9.3576 & $K_8$ &  -9.643\\
$\Gamma_8^-$ &  -7.5715 & $A_8$ &   -4.406 & $L_5$ & -7.6756 & $M_5$ &  -8.1319 & $H_8$ & -7.8786 & $K_8$ & -7.0424\\
$\Gamma_8^+$ &  -1.7207 & $A_7$ &  -4.3857 & $L_5$ & -7.6573 & $M_5$ &   -6.321 & $H_7$ & -7.8319 & $K_7$ & -7.0031\\
$\Gamma_9^+$ &  -1.6793 & $A_7$ &  -1.0552 & $L_5$ & -2.5751 & $M_5$ &  -4.7847 & $H_9$ & -4.6241 & $K_9$ & -5.3748\\
$\Gamma_7^+$ &  -0.3496 & $A_9$ &  -1.0322 & $L_5$ & -2.5516 & $M_5$ &  -3.2709 & $H_7$ & -4.5975 & $K_8$ & -3.6509\\
$\Gamma_7^-$ &  -0.0279 & $A_8$ &  -1.0192 & $L_5$ & -2.5085 & $M_5$ &    -3.01 & $H_8$ & -2.0413 & $K_9$ & -3.6304\\
$\Gamma_9^-$ &        0 & $A_9$ &  -0.9962 & $L_5$ & -2.4906 & $M_5$ &  -1.4971 & $H_9$ & -2.0303 & $K_7$ & -3.3739\\
$\Gamma_8^-$ &   1.4814 & $A_8$ &   2.3662 & $L_5$ &  1.2958 & $M_5$ &   0.7957 & $H_9$ &  2.3229 & $K_8$ &  1.2191\\
$\Gamma_9^-$ &   2.6962 & $A_7$ &   2.3878 & $L_5$ &  1.3036 & $M_5$ &   2.9848 & $H_8$ &  2.3319 & $K_9$ &  5.7005\\
$\Gamma_9^-$ &   2.7151 & $A_8$ &   3.7479 & $L_5$ &  2.4036 & $M_5$ &   3.4078 & $H_9$ &  4.1826 & $K_9$ &  6.4501\\
$\Gamma_7^+$ &   3.4206 & $A_9$ &   3.7614 & $L_5$ &  2.4125 & $M_5$ &   3.8669 & $H_8$ &  4.1891 & $K_9$ &   6.475\\
$\Gamma_7^+$ &   4.2135 & $A_7$ &   3.7624 & $L_5$ &   6.695 & $M_5$ &   5.8173 & $H_9$ &  6.8291 & $K_7$ &  6.8974\\
$\Gamma_9^+$ &   5.1671 & $A_9$ &   3.7759 & $L_5$ &   6.705 & $M_5$ &   7.2281 & $H_8$ &  6.8401 & $K_7$ &   6.919\\
$\Gamma_8^+$ &   5.1779 & $A_8$ &   7.7044 & $L_5$ &  9.1484 & $M_5$ &   7.6228 & $H_7$ &  7.3607 & $K_8$ &  6.9399\\
\hline
\end{tabular}
\caption{Transition energies at various high symmetry points and the respective irreducible representations(IR) for the lonsdaleite phase of Si. Note that the conduction band minima is very close to the $M$ valley }
\label{tab:WZ-Si}
\end{table*}

\begin{table*}
\begin{tabular}{|c|c|c|c|c|c|c|c|c|c|c|c|}
\hline\hline
\textbf{IR}&\textbf{E (eV)}&\textbf{IR}&\textbf{E (eV)}&\textbf{IR}&\textbf{E (eV)}&\textbf{IR}&\textbf{E (eV)}&\textbf{IR}&\textbf{E (eV)}&\textbf{IR}&\textbf{E (eV)}\\\hline
$\Gamma_7^+$ & -12.4692 & $A_7$ & -11.7515 & $L_5$ & -9.9534 & $M_5$ & -10.5753 & $H_7$ & -9.5329 & $K_7$ & -9.7671\\
$\Gamma_8^-$ & -10.4783 & $A_8$ &  -11.751 & $L_5$ & -9.9492 & $M_5$ &  -8.9518 & $H_8$ & -9.4188 & $K_8$ & -9.5896\\
$\Gamma_8^-$ &  -7.3386 & $A_8$ &  -4.6049 & $L_5$ & -7.6576 & $M_5$ &   -8.431 & $H_8$ & -8.0632 & $K_8$ &  -7.471\\
$\Gamma_9^+$ &  -1.7267 & $A_7$ &  -4.5941 & $L_5$ & -7.6503 & $M_5$ &   -6.548 & $H_7$ & -7.8404 & $K_7$ & -7.3368\\
$\Gamma_8^+$ &  -1.4617 & $A_9$ &  -1.1146 & $L_5$ & -2.4754 & $M_5$ &  -4.3358 & $H_7$ & -4.1686 & $K_8$ &  -4.787\\
$\Gamma_7^+$ &  -0.4896 & $A_8$ &  -1.1069 & $L_5$ & -2.4673 & $M_5$ &   -3.023 & $H_9$ & -4.0578 & $K_9$ & -3.3445\\
$\Gamma_7^+$ &  -0.1293 & $A_7$ &  -0.8801 & $L_5$ & -2.3285 & $M_5$ &  -2.6832 & $H_9$ & -1.9988 & $K_7$ & -3.3327\\
$\Gamma_9^-$ &        0 & $A_9$ &  -0.8723 & $L_5$ & -2.3199 & $M_5$ &  -1.4302 & $H_8$ & -1.9369 & $K_9$ &  -2.951\\
$\Gamma_8^-$ &   0.3103 & $A_8$ &   1.0951 & $L_5$ &  0.7849 & $M_5$ &   0.6828 & $H_7$ &  1.7489 & $K_8$ &  1.5267\\
$\Gamma_7^+$ &   0.7659 & $A_7$ &   1.1023 & $L_5$ &  0.7877 & $M_5$ &   2.0997 & $H_9$ &  1.8076 & $K_7$ &  5.5903\\
$\Gamma_7^+$ &   2.5368 & $A_7$ &    3.538 & $L_5$ &  2.4102 & $M_5$ &   3.3933 & $H_8$ &  4.1879 & $K_9$ &  5.6603\\
$\Gamma_9^-$ &   2.6714 & $A_9$ &   3.5433 & $L_5$ &  2.4151 & $M_5$ &   3.7561 & $H_7$ &  4.2033 & $K_7$ &  5.6979\\
$\Gamma_7^+$ &   3.3102 & $A_7$ &   3.6279 & $L_5$ &  6.3148 & $M_5$ &   5.0137 & $H_7$ &  6.5626 & $K_7$ &  5.7016\\
$\Gamma_8^-$ &   4.7797 & $A_9$ &   3.6332 & $L_5$ &  6.3199 & $M_5$ &   6.6232 & $H_7$ &  6.6076 & $K_7$ &  5.7153\\
$\Gamma_9^+$ &   4.8536 & $A_7$ &   6.0862 & $L_5$ &  8.3515 & $M_5$ &    6.947 & $H_9$ &  6.6965 & $K_8$ &  6.4529\\
\hline
\end{tabular}
\caption{Transition energies at various high symmetry points and the respective irreducible representations(IR) for the lonsdaleite phase of Ge. This is a direct gap semiconductor with a band gap of $0.310$ eV.}
\label{tab:WZ-Ge}
\end{table*}

\begin{table}
\center
\begin{tabular}{|c | c||c | c||c | c|}
\hline\hline
\multicolumn{2}{|c||}{\bf diamond} & \multicolumn{2}{c||}{\bf Si} & \multicolumn{2}{c|}{\bf Ge}  \\
\hline
$m_{||}$	&	$m_{\perp}$		&   $m_{||}$	&	$m_{\perp}$		&   $m_{||}$	&	$m_{\perp}$ \\
\hline
1.1376	&	1.1947	&	1.1365	&	1.1522	&	1.1887	&	1.1988 \\
0.1915	&	1.1724	&	0.2697	&	1.1129	&	0.3375	&	1.1500 \\
0.1482	&	1.8260	&	0.184	&	2.4867	&	0.2153	&	11.7707 \\
1.1656	&	0.3104	&	1.3479	&	0.1843	&	1.347	&	0.1550 \\
1.1697	&	0.3425	&	1.3598	&	0.1916	&	1.4218	&	0.1404 \\
0.1810	&	0.6821	&	0.1028	&	0.7687	&	0.0587	&	0.3416 \\
0.3414	&	0.2437	&	0.5481	&	0.201	&	0.1484	&	0.0871 \\
0.3418	&	0.3226	&	0.5637	&	0.2128	&	0.6035	&	0.0672 \\
0.7860	&	0.3785	&	1.0483	&	0.1224	&	1.0563	&	0.0852 \\
0.7877	&	0.3588	&	0.6496	&	1.096	&	0.0516	&	0.0410 \\
0.2504	&	0.9981	&	0.6548	&	0.9096	&	0.6213	&	1.0951 \\
0.2504	&	2.6106	&	0.1527	&	1.0882	&	0.6604	&	1.5539 \\
0.9974	&	0.5130	&	4.0888	&	0.1375	&	4.5651	&	0.4412 \\
\hline
\end{tabular}
\caption{Zone center effective masses, parallel and perpendicular to the $c$ axis, for diamond, Si and Ge.}
\label{tab:WZ-mass}
\end{table}

\begin{table}[h]
\begin{center}
\begin{tabular}{c|c|c|c|c}
\hline \hline
     & $E_g$ (eV) & $\Delta_{so}$ (eV) & $\Delta_{cr}$ (eV) & $\Delta{E}_{VB}$ (eV) \\
\hline
       diamond &      $4.7672~(K_9)$ &         0.0045 &       1.6950 &     0.0286  \\
            Si &      $0.7957~(M_5)$ &          0.044 &       0.3336 &    -0.1484 \\
            Ge &      $0.310~(\Gamma_8)$ &      0.404 &       0.1730 &    -0.1454 \\
\hline
\end{tabular}
\caption{ Energies of the lonsdaleite phase of diamond, Si and Ge. The symmetry of the conduction band minimum is indicated in parenthesis along with the band gap. $\Delta_{so}$ and $\Delta_{cr}$ are the spin-orbit splitting and crystal-field splitting energies extracted using Eq.~(\ref{WZ:CrSo}). $\Delta{E}_{VB}=E_{VB}^{cubic}-E_{VB}^{LD}$, is the energy difference between the top of the valance bands for the two polytypes.}
\label{tab:WZ-compact}
\end{center}
\end{table}

\section{Optical Properties} \label{sec:optical}

\subsection{Calculations} \label{sec:optical:method}
At normal incidence, for any given  polarization, the reflectivity acquires the simple form $R=\left|(1-n_i)/(1+n_i)\right|^2$, where, $n_i$ ($i=x,y$ or $z$ depending on the surface normal) is the complex index of refraction. In the linear response regime, $n(\omega)=\sqrt{\epsilon(\omega)}$ and the complex dielectric function can be separated into real and imaginary parts $\epsilon(\omega)=\epsilon^{\prime}(\omega)+i\epsilon^{\prime\prime}(\omega)$, which are related to each other by the Kramers-Kronig relations.

All dielectric function calculations in this paper are carried out in the long wavelength limit, ({\it i.e.} assuming only direct band-to-band transitions (same $k$). We obtain $\epsilon^{\prime\prime}(\omega)$ using our empirical pseudopotential wave functions.
In the electric dipole approximation, the direct transition between an initial state, $I$, and a final state, $J$, $\epsilon^{\prime\prime}(\omega)$ is given by
\begin{eqnarray}\label{eps2}
\epsilon^{\prime\prime}(\omega)&=& \left(\frac{\hbar\pi^2e^2}{m^2\omega^2}\right)\times\\\notag
 &~&\displaystyle\sum_{ij}\displaystyle{\int_{BZ}}|M_{IJ}|^2\delta(E_{c,j}({\bf k})-E_{v,i}({\bf k})-\hbar\omega)d^3k
\end{eqnarray}
where $\int_{BZ}$ is an integration over the entire Brillouin zone (BZ), $\sum_{ij}$ is a sum over all initial valance band and final conduction band states, and $E_v(\bf k)$ and $E_c(\bf k)$ are the valance and conduction band energies at their respective $\bf k$s.
The delta function is approximated by
\begin{eqnarray}
\delta(\Delta E -\hbar\omega)\approx {2}\left(1+ \cosh[\gamma(\Delta E -\hbar\omega)]\right)^{-1}
\label{delta}
\end{eqnarray}
where $\gamma$ is an adjustable damping parameter that can be used to phenomenologically incorporate lifetime broadening effects. We used $\gamma=100~\rm eV^{-1}$ which gives a transition linewidth of about $35~\rm meV$ \cite{Wang1981}.

The real part of the dielectric function, $\epsilon^{\prime}(\omega)$ is then obtained using the Kramers-Kronig relations
\begin{eqnarray}
\epsilon^{\prime}(\omega)=1+\frac{2}{\pi} {\mathcal P}\displaystyle\int_0^\infty\frac{\omega'\epsilon^{\prime\prime}(\omega')}{\omega'^2-\omega^2}d\omega'
\label{eps1_KK}
\end{eqnarray}
where $ {\mathcal P}$ is the Cauchy principle value.

The momentum matrix elements, for band to band transitions, between the initial state $I$ and and final state $J$ is given by
\begin{eqnarray}
M_{IJ}({\bf k})=\langle\phi_{I,\bf k}|\hat{p}|\phi_{J,\bf k}\rangle
\label{Mij_raw}
\end{eqnarray}
where, $\hat p$ is the momentum operator. We calculate $M_{IJ}$  using the pseudopotential wave functions, which can be written as

\begin{eqnarray}
\phi_{I,\bf k}({\bf r})=\displaystyle\sum_{\bf G}c_I({\bf k,G})\exp[i{\bf (k+G)}\cdot{\bf r}]
\label{P_psi}
\end{eqnarray}
where, $c_I({\bf k,G})$ are the eigenvector coefficients, for the $I^{th}$ state, obtained by diagonalizing the pseudopotential Hamiltonian at a given wavevector $\bf k$. Using Eq.\ref{P_psi}, $M_{IJ}$ can therefore be rewritten in terms of the expansion coefficients as follows,

\begin{eqnarray}
M_{IJ}({\bf k})=i\displaystyle{\sum_{\bf G}}c_I^*({\bf k,G})c_J({\bf k,G})[\bf{(k+G)}\cdot{\bf\hat e}]
\label{Mij}
\end{eqnarray}
where, $\bf\hat e$ is the polarization vector. The frequency dependent imaginary part of the dielectric function, $\epsilon^{\prime\prime}(\omega)$, is then be calculated using Eq.\ref{Mij}, for light polarized parallel and perpendicular to the $c$-axis. In order to evaluate the Brillouin zone integral in Eq.\ref{eps2}, we have used a set of $4.5\times 10^4$  special $k$ points. These special $k$ points are generated using the scheme of Monkhorst and Pack \cite{Monkhorst1976}

The momentum matrix elements calculated using the pseudopotential wave functions need to be corrected for the missing core states \cite{Heine.book}.
One way to do this is to include the commutator of the nonlocal pseudopotential and the position operator \cite{Read1991,Adolph1996}, while other proposed methods involve the inclusion of a core repair term \cite{Kageshima1997}. However it should be noted that both techniques cause small changes to the dielectric function (typically less than $5\%$). Some first principles calculations have even shown that there is almost no difference between dielectric functions calculated using {\it ab initio} pseudopotential wave functions and those calculated using true electron wave functions\cite{Monachesi2001}.

Our efforts to calculate the dielectric functions of the LD phase of group-IV semiconductors however greatly aided by the fact that the cubic phase dielectric functions of diamond, Si and Ge are already well known. We take advantage of the fact that the pseudopotentials are being transferred between polytypes. Due to the similarity of the electronic environment around an atom in the two polytypes, corrections to account for the missing core states should be nearly the same and transferable between polytypes. This method also accounts for local field effects and static screening effects.

First the dielectric functions for cubic diamond, Si and Ge are evaluated. We then normalize the calculated $\epsilon(\omega=0)$ to the experimentally known static dielectric constant by making use of the optical sum rule
\begin{equation}
  \epsilon_o=1+\kappa\frac{2}{\pi}\displaystyle\int_0^\infty\frac{\epsilon^{\prime\prime}(\omega)d\omega}{\omega}
\label{f-sum}
\end{equation}
\noindent
where $\kappa$ is a scaling constant which is adjusted so that the calculated $\epsilon_0$ matches experimental results from Ref. \onlinecite{Madelung}). Next, the LD phase dielectric functions are calculated based on their respective band structures. The final LD phase $\epsilon^{\prime\prime}_{\perp}$ and $\epsilon^{\prime\prime}_{||}$ are obtained by appropriately scaling their respective static dielectric functions $\epsilon_o^{\perp}$ and $\epsilon_o^{||}$. This scaled $\epsilon_o^{\perp}$ and $\epsilon_o^{||}$ is obtained using the $f-$sum rule (Eq. \ref{f-sum}) and the constant $C$ (which is already determined from the cubic phase calculations).

The static dielectric function needs to be corrected  for the missing core states and  other effects such as the influence of local fields and static screening. The local field effects often  arise from the inhomogeneity of the crystal\cite{Adler1962,Wiser1963} 
and are difficult to account for in dielectric function calculations. However as they are known to alter the static dielectric function\cite{} 
our scaling method includes the local field effects as well. This is further justified by the fact that the similarity of the local electronic environment in both polytypes.


\subsection{Predicted Dielectric Functions} \label{sec:optical:results}

\begin{figure}
\centering
\includegraphics[width=1\columnwidth]{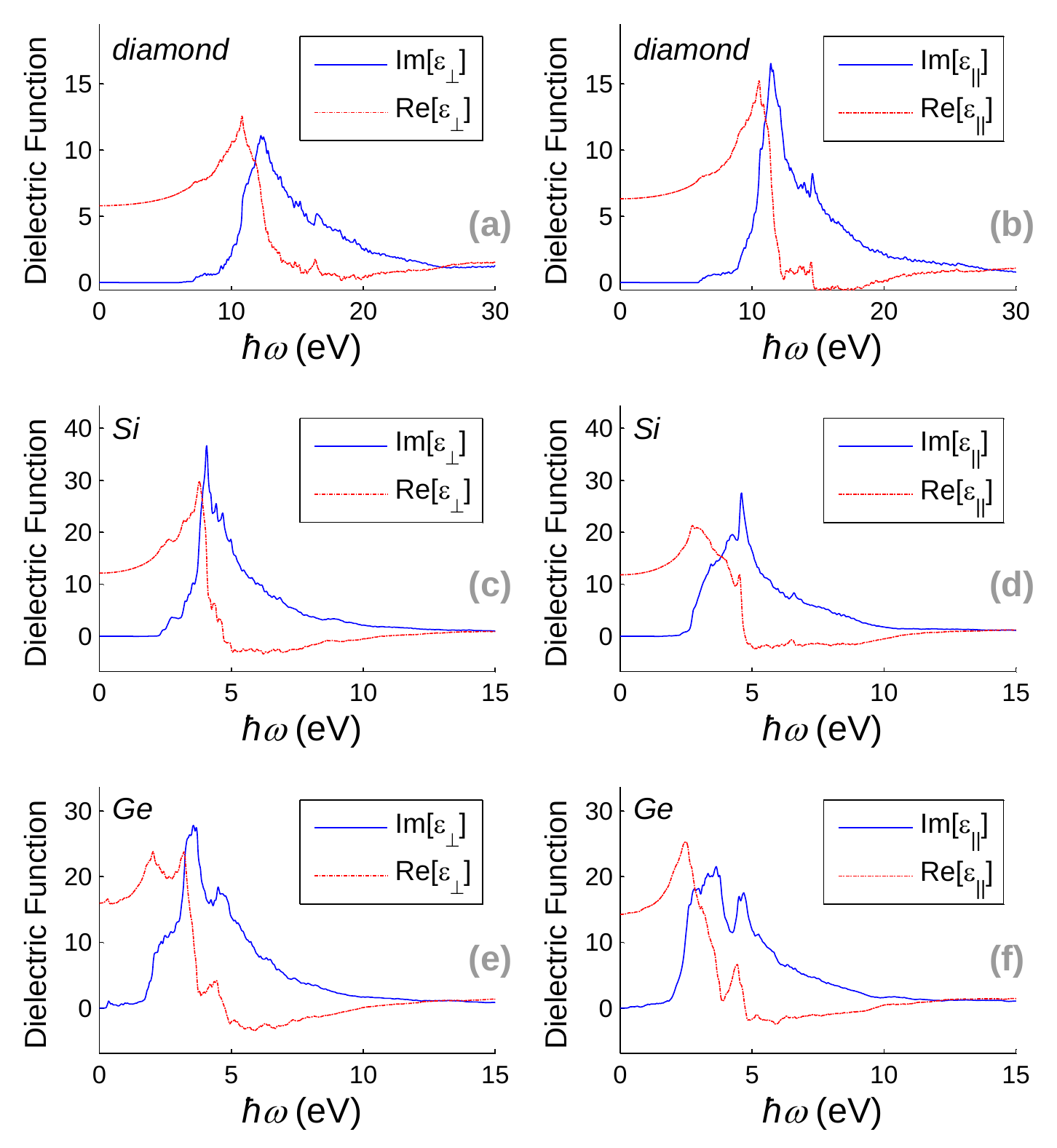}
\caption{Real and imaginary parts of the complex dielectric function as a function of optical frequency shown here for light polarized   {\bf(a)}$E_{\perp}$ in lonsdaleite phase diamond, {\bf(b)}$E_{||}$ in lonsdaleite phase diamond, {\bf(c)}$E_{\perp}$ in lonsdaleite phase Si, {\bf(d)} $E_{||}$ in lonsdaleite phase Si, {\bf(e)}$E_{\perp}$ in lonsdaleite phase Ge and {\bf(f)}$E_{||}$ in lonsdaleite phase Ge }
\label{fig:eps}
\end{figure}

\begin{figure}
\centering
\includegraphics[width=0.75\columnwidth]{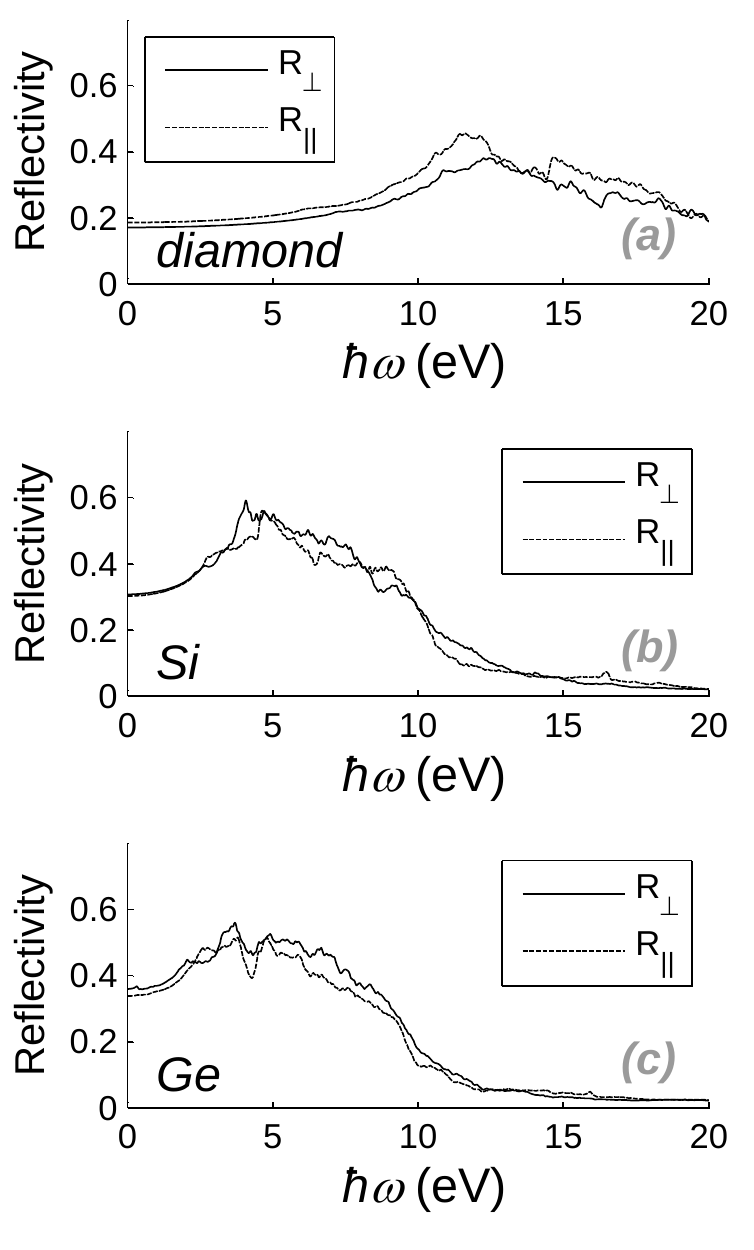}
\caption{ Calculated reflectivity spectra at normal incidence for the lonsdaleite phase of {\bf(a)} diamond, {\bf(b)} Si and {\bf(c)} Ge }
\label{fig:R}
\end{figure}

\begin{sidewaystable}
{\footnotesize
\begin{tabular}{|c|c|c c c c c|c c c c c|c c c c c|c c c c c|}
\cline{3-22}
\multicolumn{2}{c|}{} & \multicolumn{5}{c|}{for $\epsilon^{\prime}_{\perp}$} & \multicolumn{5}{c|}{for $\epsilon^{\prime\prime}_{\perp}$} & \multicolumn{5}{c|}{for $\epsilon^{\prime}_{||}$} & \multicolumn{5}{c|}{for $\epsilon^{\prime\prime}_{||}$} \\
\hline
material & $\hbar\omega_K$ (eV) & $f$ & $\Omega_1$ & $\Gamma_1$ & $\Omega_2$ & $\Gamma_2$ & $f$ & $\Omega_1$ & $\Gamma_1$ & $\Omega_2$ & $\Gamma_2$ & $f$ & $\Omega_1$ & $\Gamma_1$ & $\Omega_2$ & $\Gamma_2$& $f$ & $\Omega_1$ & $\Gamma_1$ & $\Omega_2$ & $\Gamma_2$\\\hline
diamond &10.48&353.16&12.784&10.25&11.648&1.8333&72.862&5.245&-7.51$\times 10^{-6}$&10.77&1.9892&597.98&22.364&23.007&12.046&2.0037&142.12&5.245&-6.88$\times 10^{-6}$&10.796&1.8469\\
Si &3.12&104.02&5.4846&0.0009&3.6958&-1.2413&6.9069&1.555&-3.87$\times 10^{-5}$&2.9028&0.5947&98.144&6.3892&17.455&3.4005&0.8451&12.497&1.525&-2.14$\times 10^{-5}$&3.0548&0.4527\\
Ge &2.41&102.04&8.1094&108.25&2.7345&-1.0634&14.305&1.1317&4.76$\times 10^{-6}$&2.3213&0.6054&83.872&5.0129&10.645&2.8946&-0.7297&15.061&1.1947&1.85$\times 10^{-5}$&2.5826&0.584\\
\hline
\end{tabular}
}
\caption{Lorentz Oscillator model fit coefficients for the dielectric functions of the lonsdaleite phase of diamond, Si and Ge, obtained about their respective fundamental absorption edges. These fits, for the dispersion relations, are valid only up to a cut off of $\hbar\omega_K$ as listed in the table.}
\label{tab:lorentz}
\end{sidewaystable}

As mentioned earlier, all zone center states belong to either $\Gamma_7$, $\Gamma_8$ or $\Gamma_9$ representations. The allowed interband transitions for LD zone center states are as follows \cite{Birman1959b,Streitwolf1969}. For light polarized parallel to the $c$-axis ($E_{||}$), the optical selection rules only allow transitions between same symmetry states, {\it i.e.} $\Gamma_7 \leftrightarrow \Gamma_7$, $\Gamma_8 \leftrightarrow \Gamma_8$ and $\Gamma_9 \leftrightarrow \Gamma_9$.  For light polarized perpendicular to the $c$-axis($E_{\perp}$), the allowed transitions are $\Gamma_7 \leftrightarrow \Gamma_7$, $\Gamma_8 \leftrightarrow \Gamma_8$ , $\Gamma_9 \leftrightarrow \Gamma_7$ and $\Gamma_9 \leftrightarrow \Gamma_8$. Note, that the $\Gamma_7 \leftrightarrow \Gamma_8$ transition is forbidden for all polarizations.

The other allowed high symmetry point transitions for $E_{||}$ are $A_{7,8}\leftrightarrow A_{7,8}$, $A_9\leftrightarrow A_9$, $K_{4,5}\leftrightarrow K_{4,5}$, $K_{6}\leftrightarrow K_{6}$, $H_{4,5}\leftrightarrow H_{4,5}$ and $H_{6}\leftrightarrow H_{6}$.
 Whereas for $E_{\perp}$ the allowed transitions are $A_{7,8}\leftrightarrow A_{7,8}$, $A_9\leftrightarrow A_{7,8}$, $K_{4,5}\leftrightarrow K_{6}$ and $H_{4,5}\leftrightarrow H_{6}$. All M and L valley transitions are allowed for all polarizations ({\it i.e.} $M_{5}\leftrightarrow M_{5}$ and $L_{5}\leftrightarrow L_{5}$)\cite{Streitwolf1969,Mojumder1982}.

In our calculations, Ge is the only group IV semiconductor that has a direct band gap in the LD phase. Its conduction band minima has $\Gamma_8$ symmetry while the top of the valance band has $\Gamma_9$ symmetry and hence LD phase Ge will only be optically active for $E_{\perp}$. For $E_{||}$ only transitions from deep within the valance band will be allowed. The LD phase diamond and Si on the other hand are indirect gap semiconductors and will not be optically active.

We calculate the frequency dependent $\epsilon^{\prime}(\omega)$ and $\epsilon^{\prime\prime}(\omega)$ for parallel and perpendicularly polarized light for diamond, Si and Ge as shown in Fig.~\ref{fig:eps}. As seen in Fig.\ref{fig:R}, the corresponding reflectivity spectra for both polarizations shows several peaks which originate from interband transitions along various high symmetry points. Each of these crystals have distinct spectral features depend on the details of their electronic structure. For Ge and Si the most prominent features are typically seen up to about $4~\rm eV$, whereas for diamond the most prominent peak is seen at about 12 eV. All the group-IV LD phase semiconductors are optically anisotropic as expected. However, diamond exhibits greater optical anisotropy than Si or Ge as seen in the reflectivity spectra (Fig.\ref{fig:R}). Notice that around 12 eV, the absorption is significantly more for $E_{\perp}$ due the much smaller $\epsilon_{\perp}$ structures at the same frequency.

In order to elucidate the the dielectric function's variations about the fundamental absorption edge (FAE), we fit the numerically calculated $\epsilon_{\perp}$ and $\epsilon_{||}$, bellow the Restrahlen band, to a classical Lorentz oscillator, whose real and imaginary parts are
\begin{eqnarray}
\epsilon^{\prime}(\omega)=1-f\displaystyle\sum_{j=1}^{2}\frac{\omega^2-\Omega_j^2}{(\omega^2-\Omega_j^2)^2+(\Gamma_j\omega)^2}
\label{Lorentz_eps1}
\end{eqnarray}
\begin{eqnarray}
\epsilon^{\prime\prime}(\omega)=f\displaystyle\sum_{j=1}^{2}\frac{\Gamma_j\omega}{(\omega^2-\Omega_j^2)^2+(\Gamma_j\omega)^2}.
\label{Lorentz_eps2}
\end{eqnarray}
Here, $f,~\Omega_j$ and $\Gamma_j$ were used as fitting parameters and are listed in Table \ref{tab:lorentz}. Typically, $f$ represents the oscillator strength, $\Gamma_j$ the relaxation rate and $\Omega_j$ is a resonance frequency term.

These fits in terms of the Lorentz oscillators provides an analytical expression for the  dielectric function's dispersion relations over the spectral region where there are no discontinuities. Therefore these dispersion relations are valid only within a certain cut-off frequency ($\omega_K$) as listed in Table \ref{tab:lorentz}). The fitting parameters, such as $f$s, $\Gamma_i$s and $\Omega_i$s, depend on $\omega_K$ and do not necessarily represent trends in the optical properties of these semiconductors. These analytic dispersion relations could be useful for modeling optical devices and multi-layer thin-film structures.

\begin{table}
\begin{center}
\begin{tabular}{c|c c}
 \hline
 \hline
{Material}  & $\epsilon_{0}^{\perp}$ & $\epsilon_{0}^{||}$ \\
\hline
    diamond  &  5.7988  & 6.3192  \\
    Si       &  12.1808 &  11.8518 \\
    Ge       &  16.0066 &  14.2896 \\

\hline
\end{tabular}
\caption {Calculated static dielectric constant for light polarized parallel and perpendicular to the $c$-axis, for the three lonsdaleite phase group IV semiconductors. \label{tab:eps0} }
\end{center}
\end{table}

The static dielectric constants were calculated from the imaginary part of the dielectric functions $\left(\epsilon_o=1+2\int_0^\infty \epsilon^{\prime\prime}(\omega)d\omega/{\pi\omega} \right)$, for light polarized parallel and perpendicular to the $c$ axis, and are listed in Table \ref{tab:eps0}. In general, the semiconductor with the higher atomic number(Z) has the larger static dielectric constant. We also see that $\epsilon_{0}^{\perp} < \epsilon_{0}^{||}$ in the case of diamond (which has a $\Gamma_7$ direct gap), whereas  $\epsilon_{0}^{\perp} > \epsilon_{0}^{||}$ for Si and Ge (which have $\Gamma_8$ direct gaps).

A simple explanation for this can be provided based on the optical selection rules. For a given material, $\epsilon_o$ depends on the number of allowed transitions and the oscillator strength of each transition. The closer the bands are, the stronger the dipole transitions will be. Consider a small region of $\omega$ around the direct band gap where the transitions are the strongest for zone center states. For $E_{\perp}$ a transition between the $\Gamma_9$ heavy-hole (HH) and the $\Gamma_7$ (or $\Gamma_8$) conduction band is allowed. All three materials Si, Ge and diamond will be optically active. Whereas for $E_{||}$ the $\Gamma_7^{(1,2)}$ light-hole/split-off hole to $\Gamma_7$ conduction band is only allowed for diamond. Si and Ge are completely optically dark in this case. Hence for $E_{||}$, diamond has more allowed transitions than for $E_{\perp}$ (assuming similar oscillator strengths) and therefore its $\epsilon_{0}^{\perp} < \epsilon_{0}^{||}$. Whereas the opposite is true for Si and Ge.

\section{Summary}\label{sec:summary}
We have calculated the electronic band structures and  dielectric functions for diamond, Si and Ge in lonsdaleite phase using transferable model potentials, including spin-orbit coupling. The potentials should be accurate since the local electronic environment for the cubic and lonsdaleite polytypes are very similar. It is seen that while diamond and Si remain indirect  in the lonsdaleite phase, Ge is transformed into a narrow direct-gap semiconductor due to zone folding effects. Hence LD-type Ge will be optically active, which could make it extremely useful for technological applications. We have also tabulated a number of parameters such as high symmetry point energies, their irreducible representations and effective masses, which could be useful for constructing $k\cdot p$ type models. We calculated the frequency dependent complex dielectric functions up to $20~\rm eV$ for light polarized parallel and perpendicular to the $c$-axis in the dipole approximation.
We find strong optical anisotropy, making LD phase materials potentially useful as nonlinear crystals since their optical birefringence enables them to satisfy phase matching conditions.

\section{Appendix}\label{sec:appendix}

\begin{figure}
\includegraphics[width=0.8\columnwidth]{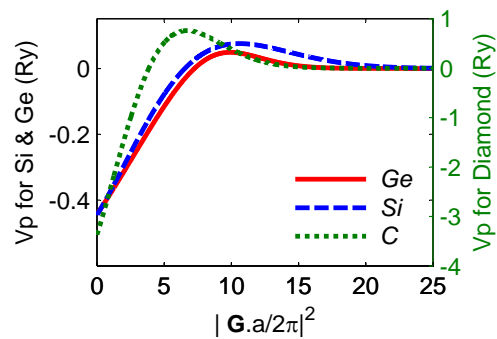}
\caption{Local atomic form factors($V_p$) for diamond(C), Si and Ge. Note that the y-axis for diamond is on the right.}
\label{fig:PP}
\end{figure}

\begin{figure}
  \includegraphics[width=1.0\columnwidth]{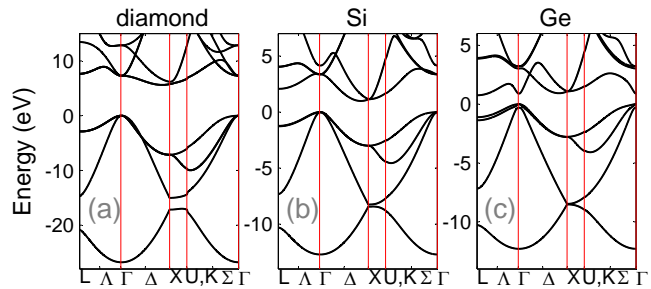}
  \caption{ Calculated band structures, with the inclusion of spin-orbit interactions, for the cubic phase of (a) diamond (b) Si and (c) Ge }
  \label{fig:ZB}
\end{figure}

\begin{table*}
\center
\begin{tabular}
{c | c | c || c |c |c || c |c |c}
\hline \hline
\multicolumn{3}{ c||}{\bf diamond} & \multicolumn{3}{c||}{\bf Si} & \multicolumn{3}{c }{\bf Ge} \\
\hline
     $Transiton$ &  $Expt.~(eV)$ &  $Calc.~(eV)$ & $Transition$ &  $Expt.~(eV)$ &  $Calc.~(eV)$ & $Transition$ &  $Expt.~(eV)$ &  $Calc.~(eV)$  \\
\hline
$\Gamma_{7c}^--\Gamma_{8v}^+$ & 7.3  & 7.3    &  $\Gamma_{7c}^--\Gamma_{8v}^+$ & 3.35 & 3.35   & $\Gamma_{7c}^--\Gamma_{8v}^+$ & 0.898 &      0.898\\
$\Gamma_{6c}^--\Gamma_{8v}^+$ & 12.9 & 12.854 &  $\Gamma_{6c}^--\Gamma_{8v}^+$ & 4.15 & 4.148  & $\Gamma_{6c}^--\Gamma_{8v}^+$ & 3.22  &      3.223\\
$\Gamma_{8v}^+-\Gamma_{6v}^+$ &  26  & 26.779 &  $\Gamma_{8v}^+-\Gamma_{6v}^+$ & 12.5 & 12.707 & $\Gamma_{8v}^+-\Gamma_{6v}^+$ & 12.6  &      12.286\\
$L_{6c}^+-L_{6v}^-$           & 10.5 & 10.508 &  $L_{6c}^+-\Gamma_8^+$         & 2.05 & 2.072  & $L_{6c}^+-\Gamma_8^+$         & 0.76  &      0.76  \\
 $X_{5c}-X_{5v}$              & 12.9 & 12.89  &  $X_{5c}-\Gamma_8^+$           & 1.13 & 1.13   & $X_{5c}-\Gamma_8^+$           & 1.16  &      1.081 \\
$\Delta_{so}$                 & 0.006&  0.006 &   $\Delta_{so}$                & 0.441& 0.441  & $\Delta_{so}$                 & 0.297 &      0.297 \\
$\Delta_{so}^\prime$          &    - &      - &   $\Delta_{so}^\prime$         &    - &      - & $\Delta_{so}^{\prime}$        &   0.2 &      0.2  \\
\hline
\end{tabular}
\caption{A comparison between targeted experimental transition energies taken from Ref.[\onlinecite{Madelung}] and the calculated values for cubic diamond, Si and Ge. For each material, the first column shows the targeted transitions, the second column shows the targeted energies and the third column shows the converged results from fitting the pseudopotentials. The spin-orbit splitting energies are $\Delta_{so}=E^{\Gamma}_{8v} - E^{\Gamma}_{7v}$ and for Ge, $\Delta_{so}'=E^{\Gamma}_{8c} - E^{\Gamma}_{6c}$.}
\label{tab:ZB-conv}
\end{table*}

\begin{center}
\begin{table*}
\begin{tabular}{c|c c c c c c}
 \hline
 \hline
{Material} & $x_1$ & $x_2$  &  $x_3$  &  $x_4$ & $\mu_1$ & $\mu_2$ \\
\hline
diamond & 444.305 &  -1716.53 &  0.0263  &   6.2294   & 12.515 & - \\
Si  &  8.2808  &  -54.1842   & 0.0116  &  4.7922   &  0.0536  & - \\
Ge  &  0.0791  &  -0.5737  &  0.0247  &  -1.2269  & 0.2413  & 8.4571e-17  \\
\hline
\end{tabular}
\caption { The $x_js$ are the fitting parameters for the form factors and spin-orbit splitting energies and $\mu_1$ and $\mu_2$ are the fitting parameters for the spin-orbit coupling. Note that the form factors are in units of $Ry$.}
\label{tab:ZB-FF}
\end{table*}
\end{center}


In this appendix we provide the details for our bandstructure calculations for the cubic phase of diamond, Si and Ge. The form factors of the cubic polytypes are required for obtaining the bandstructures of the hexagonal polytypes. The Hamiltonian, given in Eq.\ref{HoVso}, is diagonalized with a plane wave basis cutoff of $|{\bf G}| \le 64 \pi / a$. As a starting point, our continuous form factors for Si and Ge were first fit to the discrete form factors of Ref.[\onlinecite{Chelikowsky1976}] and for diamond, they were fit to the discrete form factors of Ref.[\onlinecite{Saravia1968b}]. The adjustable parameters $x_i$, and $\mu_l$ in Eqs.~\ref{FF.func1} and \ref{lam} were then adjusted to fit the calculated band structure to the experimental energies of the band extrema of the cubic materials. A modified simulated annealing method was used for the fitting procedure (see Ref.\onlinecite{De2010prb} for more details).

Five adjustable parameters for diamond and Si, and a total of six adjustable parameters for Ge were used to fit the calculated bandstructure to seven different experimental transition energies (which were obtained from Ref. \onlinecite{Madelung}). Additional constraints were imposed to ensure the correct band ordering of valence states and conduction states.

As can be seen in table.\ref{tab:ZB-conv}, our fits are in excellent agreement with experiment. The fitting parameters for the form factors, used for obtaining the results shown in table.\ref{tab:ZB-conv}, are given in table.\ref{tab:ZB-FF} and the atomic form factors themselves are shown in fig.\ref{fig:PP}. The calculated ZB bandstructures of these group-IV semiconductors are shown in fig.\ref{fig:ZB}. Note that for cubic diamond and Si, the $p$-like conduction band states are bellow the $s$-like state. In addition, these are also indirect gap semiconductors with the conduction band minima lying in very close proximity to the $X$-valley along the $\Delta$-direction.




\end{document}